\documentclass[prl,twocolumn,showpacs,preprintnumbers,amsmath,amssymb]{revtex4}

\usepackage{graphicx}% Include figure files
\usepackage{dcolumn}% Align table columns on decimal point
\usepackage{bm}% bold math

\newcommand{\comment}[1]{}

\renewcommand{\eqref}[1]{Eq.~(\ref{#1})}

\newcommand{\figref}[1]{Fig.~\ref{#1}}

\begin{document}

\title{From cellular properties to population asymptotics \\
in the Population Balance Equation}

\author{Tamar Friedlander$^1$}
\author{Naama Brenner$^2$}

\affiliation{$^1$Department of Physics, $^2$Department of Chemical
Engineering, Technion-Israel Institute of Technology, Haifa 32000,
Israel }

\date{\today}

\begin {abstract}
Proliferating cell populations at steady state growth often exhibit
broad protein distributions with exponential tails. The sources of
this variation and its universality are of much theoretical
interest. Here we address the problem by asymptotic analysis of the
Population Balance Equation. We show that the steady state
distribution tail is determined by a combination of protein production and
cell division and is insensitive to other model details. Under general
conditions this tail is exponential with a dependence on parameters
consistent with experiment. We discuss the conditions for this
effect to be dominant over other sources of variation and the
relation to experiments.
\end{abstract}


% 87 - Biological and medical physics
% 87.10.+e General theory and mathematical aspects
% 87.15.Aa Theory and modeling; computer simulation
% 87.17.Ee Growth and division
% 87.23.Cc Population dynamics and ecological pattern formation

 %Use showkeys class option if keyword display desired
\maketitle

%%%%%%%%%%%%%%%%%%%%%%%%%%%%%%%%%%%%%%%%%%%%%%%%%%%%%%%%%%%%%%%%%%%%%%%%%%%%%%%%%%%%%%%%%%%%

%\section{Introduction}

%%%%%%%%%%%%%%%%%%%%%%%%%%%%%%%%%%%%%%%%%%%%%%%%%%%%%%%%%%%%%%%%%%%%%%%%%%%%%%%%%%%%%%%%%%%%

Biological cell populations are diverse in their physiological
properties, even if genetically identical. Since physiology rather
than genetics ultimately carries biological function, there is much
interest in understanding this aspect of biological variation. A
good model system for this problem is a microorganism population
that is genetically uniform and grows under uniform conditions;
these systems have been studied for many years, and have recently
received renewed attention following developments in experiment
design and technique of single-cell measurements (reviewed
by~\cite{Kaern_review05, Avery06}). Experiments using fluorescence
tagging combined with microscopy and cytometry have focused on
variation in particular proteins inside cells, while theoretical
studies have provided  models of specific circuits and noise
sources. Under steady state growth conditions, several experiments
have shown that even for regulated proteins, distribution shapes are
insensitive to many details and are often observed to be broad with
exponential tails \cite{BraunBrenner04,BrennerFarkash}. This calls
for a more physical perspective of the problem, raising questions
such as the universality of the resulting distributions. We here
show that an exponential tailed distribution
with the correct dependence on system parameters
follows from a
description involving a balance between deterministic protein production and
dilution at cell division if these processes satisfy reasonable
conditions. Such tails, reflecting variation in division time, are
thus expected even if stochastic fluctuations in gene expression are
negligible. The conditions for this effect to be dominant
relative to noise in protein production
are discussed.

% MODEL DESCRIPTION
%%%%%%%%%%%%%%%%%%%%
A general theoretical framework for describing population
distributions of quantities that obey a balance of growth and
division, such as cell size or protein content, is the Population
Balance Equations (PBE)~\cite{Fredrickson,Henson}. In its most
general form it can incorporate many details and multiple internal
cellular properties. We here focus on the case where the relevant
physiological property of each cell can be described by a single
variable $x$~\cite{DiekmannMetz,Mantzaris06}:
\begin{eqnarray}
\label{eq:mass_balance_density}  \lefteqn{ \frac{\partial}{\partial
t}f(t,x)+\frac{\partial}{\partial x}\left[
g(x)f(t,x)\right]=-b(x)f(t,x)}
\\  \nonumber
&+& 2\!\!\displaystyle
\int_0^1\frac{d(p)}{p}b\!\left(\frac{x}{p}\right)f\!\left(t,\frac{x}{p}\right)dp
-f(t,x)\!\!\displaystyle\int_0^\infty \!\!b(\xi)f\!(t,\xi)d\xi.
\end{eqnarray}
Here $f(x)$ is the probability density for the quantity $x$ in the
population, and $g(x)$ is the individual growth rate of $x$. Cell
division is assumed to follow a "sloppy control"
mechanism~\cite{LordWheals}: $b(x)$ is the probability per unit time
for a cell of quantity $x$ to divide. Once division occurs, $d(p)$
is the probability for dividing into two daughter cells with
fractions $p$ and $1\!-\!p$ of the mother cell. To obey mass
conservation $d(p)\!=\!d(1-p)$. The last term in the equation
accounts for normalization. Underlying this model is the assumption
that the growth process occurs gradually and with small fluctuations
throughout the cell cycle, whereas division abruptly induces a large
change in $x$.

\begin{figure}
\includegraphics[width=7.5cm]{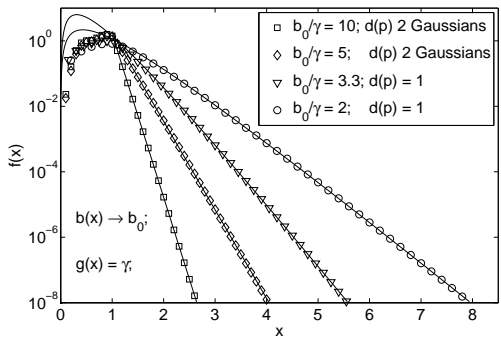}\\
\caption{Steady state population distributions with exponential
tails. Numerical solution for constant growth $g(x)\!\!=\!\!\gamma$
and saturating probability of division per unit time. $(\square,
\lozenge)$: $b(x) \!= \!b_0 H(x-\theta)$; $d(p)$ sum of two
Gaussians at $p\!=\!0.3$ and $p\!=\!0.7$. This function describes
asymmetric division, such as that observed %e.g.
for budding yeast
cells. $(\triangledown, \bigcirc)$:
 $b(x)
\!=\!\frac{b_0}{2}\left[\tanh(k(x-\theta))+1\right]$ with
 $k=5$, $\theta = 1$ and $d(p)\!=\!1$. Asymptotic approximations
 (\eqref{eq:asymp_exp}) are shown by solid lines.}
  \label{fig:exp}
\end{figure}

A large body of previous work on this model is dedicated to theorems
regarding the existence and uniqueness of solutions~\cite{Heijmans},
numerical algorithms
(\cite{Mantzaris01} and references there)
and special case solutions~\cite{Tyson86}. Traditionally the
coordinate $x$ was interpreted as related to cell size (mass, linear
dimension etc.), and the dependence of the probability per unit time
to divide on cell size reflects the combination of deterministic
size-dependent and random aspects of cell
division~\cite{LordWheals}. However, for our purpose of analyzing
the asymptotic properties of the steady-state distributions, $x$ can
also be interpreted as the amount of a particular protein or
molecule in the cell, any quantity which is produced and preserved
at cell division. This follows because the probability per unit time
to divide generally saturates for large values of cell size, age, or
protein content, reflecting the inherent probabilistic component of
the cell cycle \cite{Nurse80}.
This point, as well as the effect of an additional stochastic
component in $g(x)$, will be further discussed below.

%%%%%%%%%%%%%%%%%%%%%%%%%%%%%%%%%%%%%%%%%%%%%%%%%%%%%%%%%%%%%%%%%%%%%%%%%%%%%%%%%%%%%%%%%
%   Asymptotic approximation
%%%%%%%%%%%%%%%%%%%%%%%%%%%%%%%%%%%%%%%%%%%%%%%%%%%%%%%%%%%%%%%%%%%%%%%%%%%%%%%%%%%%%%%%%

Our analysis begins by considering the steady state solution of
\eqref{eq:mass_balance_density}. Assuming such a solution exists,
$f(t,x)\!\!=\!\!f(x)$ and the last integral
becomes
a constant,
$\int_0^\infty b(x)f(t,x)dx\!=\!R$. This constant is the specific
growth rate of the number of cells in balanced exponential growth,
and can be viewed as a parameter in the equation. Therefore at steady
state,
%\begin{eqnarray}
%\label{eq:ss} \lefteqn{ \frac{d}{dx}\left[ g(x)f(x)\right]=-b(x)f(x)}\\
%\nonumber &+& 2\!\!
%\int_0^1\frac{d(p)}{p}b\!\left(x/p\right)f\!\left(x/p\right)dp
%-R~f(x).
%\end{eqnarray}
\begin{eqnarray}
\label{eq:ss} \lefteqn{ \frac{d}{dx}\left[ g(x)f(x)\right]=-b(x)f(x)}\\
\nonumber &+& 2\!\!
\int_0^1\frac{d(p)}{p}b\!\left(\frac{x}{p}\right)f\!\left(\frac{x}{p}\right)dp
-R~f(x).
\end{eqnarray}
Now consider the incoming flow contributing to the probability
density at large $x$, where $f(x)$ is a decreasing function. It
comes from two processes: growth, bringing cells of low $x$ to a
higher one; and division, breaking high-$x$ cells into pairs of
smaller $x$. If the probability density decreases rapidly enough,
then for large $x$ the first of these incoming flows is dominant
over the second. We shall assume that this is the case for now,
neglect the integral term representing the second flow in
\eqref{eq:ss}, and return to examine the consistency of this
assumption later. One then obtains the following ordinary
differential equation:
\begin{equation}
\label{eq:ss_approximation}
\frac{d}{dx}\left[g(x)f(x)\right]=-(b(x)+R)f(x)
\end{equation}
%an ordinary
%differential equation
with the solution:
\begin{equation}
\label{eq:tail_ratio} f(x) = C \exp\left(-\int^{x}
\frac{b(\xi)+R+g'(\xi)}{g(\xi)}d\xi\right).
\end{equation}
A related integral was found for the case of exactly symmetric
division and a finite ranged variable~\cite{DiekmannMetz}. Here we
argue that in general under the assumption of a rapidly decreasing
$f(x)$ the ratio between two points at the tail of the distribution
is given by \eqref{eq:tail_ratio} with the limits of integration at
the two points.

% exponential tails
%%%%%%%%%%%%%%%%%%%%

If $x$ represents cell size, $g(x)$ is the growth function of the
individual cell. Experiments directly measuring this function are
not straightforward \cite{footnote_size};
%(reviewed in\cite{ReshesBJ2008});
theoretical works have mostly assumed either
linear or constant functions for simplicity.
%%% Tamar
%theoretical literature~\cite{DiekmannMetz, Mantzaris06, Heijmans,
%Mantzaris01, Tyson86}.
%%%
%Recently, the growth of individual bacteria was measured and a
%constant rate was found for large $x$~\cite{ReshesBJ2008}.
%On physical grounds it is reasonable that
%even if $g$ depends on $x$, it will saturate at large $x$.
If $x$ is interpreted as
%%% Tamar
the
%%%
amount of a protein, then a constant $g$ represents a mean rate of
protein production that is independent of the protein level.
%; fluctuations in $g$ are neglected (see discussion below).
Assuming $g(x)\!=\!\gamma$ and a saturating probability per unit
time to divide $b(x)\!\to \!b_0$ for large $x$,
\begin{equation}
f(x) \!\sim\! e^{-\kappa
x},~~~~~~~~\kappa\!=\!(b_0\!+\!R)\!/\!\gamma. \label{eq:exp_tail}
\end{equation}

%with $\kappa\!=\!(b_0\!+\!R)\!/\!\gamma$.
\noindent Returning now to the question of the validity of the naive
approximation \eqref{eq:tail_ratio}, a resulting exponential tail
hints to consistency of the approximation since the function
decreases rapidly. More precisely, we assumed that
\begin{equation}
%\label{eq:condition} 2\int_x^\infty f(\xi)b(\xi)\frac{d \xi}{\xi}
%\ll -\frac{d}{dx}[g(x)f(x)].
\label{eq:condition}
2\int_0^1\frac{d(p)}{p}b\!\left(\frac{x}{p}\right)f\!\left(\frac{x}{p}\right)dp
\ll -\frac{d}{dx}[g(x)f(x)].
\end{equation}
%\begin{equation}
%\label{eq:condition}
%2\int_0^1d(p)/p~b\!\left(x/p\right)f\!\left(x/p\right)dp
%\ll -\frac{d}{dx}[g(x)f(x)].
%\end{equation}
Substituting the above exponential one finds that this requirement
is satisfied by $x \!>>\! d_{max}\rho/\kappa^2$, where
$d_{max}\!=\!max_{p}\{d(p)\}$ and $\rho\!=\!2 b_0/\gamma$; this
defines the regions of consistency of the approximation.

The Population Balance Equation \eqref{eq:mass_balance_density} can
be solved numerically~(\cite{Mantzaris01} and references there). We
have developed a numerical procedure to solve the time-dependent
equation on a semi-infinite range based on the method of
time-evolution operators~\cite{DiekmannMetz}. \figref{fig:exp} shows
the steady state solution with functions $g,b$ that saturate at
large $x$. As predicted by the argument above, the distributions
exhibit exponential tails. Starting the dynamics from various
initial conditions always relaxed to the same steady state
distribution. An exponential tail was found for all division
functions $d(p)$, consistent with \eqref{eq:exp_tail}.

Using this observation, we proceed without much loss of generality
to a more accurate asymptotic approximation for the case
$d(p)\!\!=\!\!1$. Assuming once again $g(x)\!\!=\!\!\gamma$ and
$b(x)\!\!\to\!\! b_0$ for large $x$, \eqref{eq:ss} is equivalent, by
a change of variables and an additional differentiation, to
\begin{equation}
\label{eq:dp1ODE}  \frac{d^2 f}{dx^2}+\kappa
\frac{df}{dx}+\rho\frac{1}{x}f(x)=0.
\end{equation}
$x\!=\!\infty$ is an irregular singular point of this
equation~\cite{Bender-Orszag}. Trying a solution
$f(x)\!\!=\!\!\exp(\mu x)x^\lambda \eta(x)$ with
$\lambda\!\!\in\!\!\mathcal{R}$ and $\eta(x)$ analytical at
$x=\infty$, we obtain to leading order:
%\begin{eqnarray}
%\label{eq:asymp_exp} f(x)_{x \to \infty} \sim A x^{\beta}
%\exp{\left(-\frac{b_0+R}{\gamma}x\right)} + B x^{-\beta}
%\end{eqnarray}
\begin{eqnarray}
\label{eq:asymp_exp} f(x)_{x \to \infty} \sim C_1 x^{\rho/\kappa}
e^{-\kappa x} + C_2 x^{-\rho/\kappa}.
\end{eqnarray}
\noindent Since \eqref{eq:dp1ODE} is of second order we have two
independent solutions; however, as $0 < R\leq b_0$ it follows that
$1 \leq \rho/\kappa < 2$ and hence the mean of the second solution
diverges. This observation, while obviously not a proof of
uniqueness, supports the numerical result of relaxation to a unique
steady state distribution from many initial conditions.

An exactly solvable case occurs when $b(x)\!=\!b_0$, then % i.e. the
%probability of division is independent of $x$:
%\[f(x)=\left(\frac{2 b_0}{\gamma}\right)^2 x
%\exp\left(-\frac{2b_0}{\gamma} x\right ).\]
$f(x)=\kappa^2 x e^{-\kappa x}$. Here $R\!=\!b_0$, then
$\kappa=\rho$ so the first asymptotic function in
\eqref{eq:asymp_exp} is an exact solution; the second, $f(x)
\!\!\sim \!\!x^{-1}$, is non-normalizable. The PBE here reduces to a
model studied in~\cite{BrennerShokef}, where protein is produced at
a constant rate and cells divide with constant probability per unit
time.

We thus establish that under general conditions the steady state
distribution exhibits an exponential tail, as has been observed in
several experiments~\cite{BraunBrenner04, BrennerFarkash}. The
exponential tail is obtained neglecting variation in the source $g$,
and stems from a balance between the first-order
kinetics
of cell
division and a constant or saturating
deterministic
source. The dependence of the
exponent on parameters is such that upon increase of production,
represented by $g$, the exponential tail broadens. This is
consistent with experimental observations on protein production at
steady state in populations of yeast cells~\cite{BrennerFarkash},
and inconsistent with most models that account for population
variation by production noise.

Formally \eqref{eq:tail_ratio} indicates that the distribution tail
is determined by the ratio of the growth and division functions, not
by each of them separately. Thus, if for large $x$ these functions
do not saturate but have the same $x$-dependence, an exponential
tail will also arise. Fig. 2 shows the numerical solution for
linearly increasing $g(x),~b(x)$, supporting this prediction. While
not immediately relevant to protein production, this result
illustrates how exponential tails can arise by different growth and
division functions maintaining constant ratio. It thus supports our
analytic conclusion about how the combination of these functions
shapes the distribution tails.

\begin{figure}
  \includegraphics[width=7.5cm]{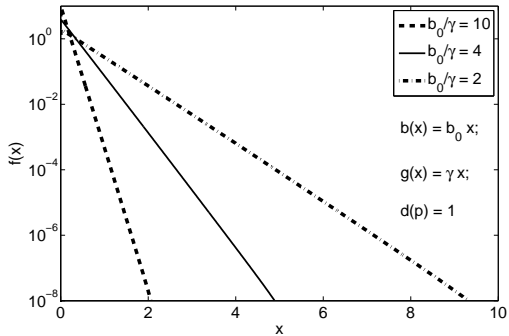} \\
   \caption{Steady state distributions for $g(x)=\gamma x$, $b(x) = b_0 x$ and $d(p)=1$.
   This example shows that the distribution tail is determined by the
  ratio between $b(x)$ to $g(x)$ and not by each function separately.}
    \label{fig:exponential_tail}
\end{figure}

\begin{figure}
  \includegraphics[width=7.5cm]{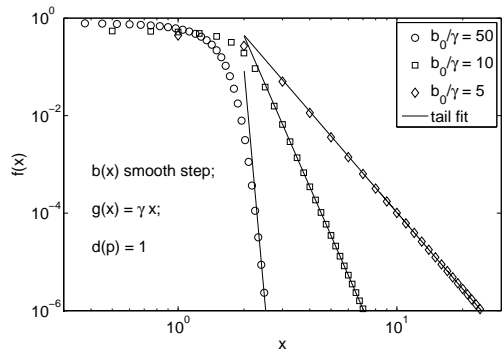} \\
   \caption{Steady state distributions for $g(x)=\gamma x$,
   $b(x)=\frac{b_0}{2}\left[\tanh(k(x-\theta))+1\right]$ with $k=2$, $\theta = 2$ and $d(p)=1$.
   Numerical
solutions
   (symbols) are shown with asymptotic approximation to the
   tail (lines).}
    \label{fig:powerlaw_tail}
\end{figure}

A growth, or production, function $g(x)$ that increases with $x$ is
relevant for several biological contexts. For example, if food
uptake is related to the surface area of the organism and $x$ is a
linear dimension, then growth is an increasing function of
$x$~\cite{DiekmannMetz}. For $g(x)\!=\!\gamma x$ one can show that
$R=\gamma$ and therefore $\kappa=\rho/2+1$.
Using the same procedure as before to
write an equivalent ordinary differential equation for
$b(x)$ saturating to $b_0$ at large $x$ and
$d(p)\!=\!1$, we find
\begin{equation}
\label{eq:asymp_power_law}
\frac{d^2f}{dx^2}+\left[\rho/2+3\right]\frac{1}{x}\frac{df}{dx}
+\rho\frac{1}{x^2}f=0.
\end{equation}
This is the Euler equation~\cite{Bender-Orszag} with power-law solutions
$f(x)=Cx^{\alpha}$ where $\alpha={-\rho/2,-2}$.
%$(\kappa+1)^2-4\rho>0$ is $f(x)= C x^{\alpha}$ with
%Here $x\!\!=\!\!\infty$ is a
%regular singular point of the equation. Applying the Frobenius
%method~\cite{Bender-Orszag} we obtain the asymptotic approximation
%$f(x)\sim x^\alpha$ with
%\begin{equation}
%\label{eq:alpha} \alpha_{1,2}=\{-(\kappa+1)\pm
%\sqrt{(\kappa+1)^2-4\rho}\}/2.
%\end{equation}
%This yields a physical solution only for $\alpha<-2$.
Of the two independent solutions to the asymptotic
equation, only $\alpha\!=\!-\rho/2$ with $\rho\!>\!4$ is consistent with $f(x)$
being a probability density with a finite mean. Indeed, numerical
simulations in this parameter regime always relax to a steady state with a
tail $f(x)\sim x^{-b_0/\gamma}$;
see \figref{fig:powerlaw_tail} for a comparison between the numerical solution
and the asymptotic tail.
%As before, the free parameter in the fit is the exponential growth
%rate $R$.

The special case of $b(x)=b_0 H(x-\theta)$, where $H$ is the
Heaviside function with threshold $\theta$, is exactly solvable.
Here the  Euler equation \eqref{eq:asymp_power_law} holds exactly in the region
$x\!>\!\theta$. By continuity and normalization requirements one can show that
the coefficient of the solution with $\alpha\!=\!-2$ is exactly zero, and the
unique solution is
%\begin{equation}
%f(x)=\left\{
%\begin{array}{lr} \frac{b_0-\gamma}{b_0} \cdot \frac{1}{\theta} & x
%\leqslant \theta
%\\
%\\
%\frac{b_0-\gamma}{b_0}\cdot %\frac{1} {\theta^{-b_0/\gamma+1}}
%\theta^{b_0/\gamma-1} x^{-b_0/\gamma} & x>\theta\end{array} \right.
%\end{equation}
\begin{equation}
f(x)=\left\{
\begin{array}{lr} (1-2/\rho) \cdot \frac{1}{\theta} & x
\leqslant \theta
\\
\\
(1-2/\rho)\cdot \theta^{\rho/2-1} x^{-\rho/2} & x>\theta\end{array}
\right.
\end{equation}

\noindent Once again, this solution is valid for
%%% Tamar
$\rho>4$ ($b_0>2\gamma$).
%$\rho>2$.
%$\gamma<b_0$.
%%%
%Since here $R=\gamma$, $2\kappa = \rho+2$, we have
%$\alpha=-b_0/\gamma$
%$\alpha=-\rho/2$ in \eqref{eq:alpha}, consistent with the exact
%solution.
Note that the naive argument leading to \eqref{eq:tail_ratio} is
self-consistent in this case only for a more severely limited region of
parameters ($b_0>>3\gamma$).

%%%%%%%%%%%%%%%%%%%%%%%%%%%%%%%%%%%%%%%%%%%%%%%%%%%%%%%%%%%%%%%%%%%%%%%%%%%%%%%%%%%%%%%%%%%%

%\section{Summary}

%%%%%%%%%%%%%%%%%%%%%%%%%%%%%%%%%%%%%%%%%%%%%%%%%%%%%%%%%%%%%%%%%%%%%%%%%%%%%%%%%%%%%%%%%%%%
In summary, we used the population balance equation (PBE) to study
the interplay between intracellular and population processes in
shaping the steady state distribution in a dividing cell population.
The novel component in our approach is to consider the variable $x$
describing the cell state as unbounded and to focus on the
asymptotic properties of its distribution. This enables us to extend
the interpretation of $x$ as a particular protein or molecule in the
cell, since asymptotically the probability per unit time to divide
becomes independent of the variable, $b(x)\to b_0$ for large $x$.
This probabilistic component of the cell cycle is a well established
property for many cell types \cite{Nurse80,LordWheals}.
%this assumption may be
%violated in a culture where aging cells, which correlate positively
%with large values of $x$, divide more slowly. It is a good
%approximation if aging cells do not accumulate, for example in
%continuous culture. This assumption makes it possible to expand the
%interpretation of $x$ as a particular protein or molecule in the
%cell. Otherwise this interpretation is still reasonable if the
%molecule correlates strongly with cell size.
%Within this interpretation, we have assumed that dissipation of the
%molecule $x$ proceeds predominantly by cell division, corresponding
%to a lifetime longer than a cell generation; direct degradation can
%be added otherwise.

We have shown that generally the functional forms of mean growth or
production $g(x)$ and probability per unit time to divide $b(x)$
determine the tail of the distribution through a particular
combination, \eqref{eq:tail_ratio}. Because the PBE takes into
account the kinetics of cell division as a discrete process,
randomness in the timing of cell division
%- a "sloppy" control of division~\cite{Tyson86} -
is sufficient to yield an exponentially tailed distribution at
steady state. In reality, the single-cell function $g(x)$ itself has
a stochastic component, and this can be added to the model using the
diffusion approximation. Such an extension will be a good
approximation if $\frac{\delta g^2}{\langle g \rangle
^2}\ll\frac{1}{b_0+R}$.

At the other extreme, if internal stochasticity is dominant, it
should be modeled in detail. For example, previous work has shown
that bursts in mRNA production cause an exponential distribution of
protein produced in each cell, which in turn is reflected as
exponential tails in the population
distribution~\cite{Berg78,Paulsson00,Friedman06}. Division can then
be assumed synchronous with symmetric binomial distribution
\cite{Berg78,Paulsson00}, or it can be altogether neglected and
described as a continuous dissipative process \cite{Friedman06},
without changing the result. The validity of each regime depends on
the relative variation of the two processes, production and
division, and on their relative time scales. One way to identify the
regime in experiment is the dependence of the exponential tail on
parameters: if the tail results from microscopic effects, then a
larger mean production results in relatively narrower distributions
and the slope of the tail remains intact. However, if the exponent
results from a combination of sloppy division and deterministic
production as suggested here, then larger mean production results in
a broader exponential tail. Experiments on yeast populations have
shown that increasing the mean protein production, either by an
increase in the number of promoters or by adding inducing agents,
increases the mean and at the same time broadens the exponential
tail~\cite{BrennerFarkash}. This dependence suggests that it is the
population effects, rather than microscopic noise, which govern the
distribution tails in these experiments.

In any interpretation of $x$, our results predict that the
distribution tails will be insensitive to the division function
$d(p)$. This is supported by the universality of protein
distribution tails in yeast cells grown under various steady state
conditions~\cite{BrennerFarkash}. Yeast cells divide asymmetrically,
with the degree of asymmetry depending on growth rate and
environment~\cite{LordWheals}. The observation that under all growth
conditions the protein distribution exhibited exponential tails is
consistent with our prediction. Moreover, unpublished results on
bacteria populations grown at steady state~\cite{Salman} show that
even this symmetrically dividing organism exhibits similar
exponential tails.

Taken together, our results suggest that exponential tails in the
distribution of an abundant protein in a dividing population may be
a much more universal feature than previously thought, since they
reflect fundamental properties of randomness in cell division times
and not necessarily the particular microscopic details of protein
production circuits.

%%%%%%%%%%%%%%%%%%%%%%%%%%%%%%%%%%%%%%%%%%%%%%%%%%%%%%%%%%%%%%%%%%%%%%%%%
% Acknowledgments:
%%%%%%%%%%%%%%%%%%%%%%%%%%%%%%%%%%%%%%%%%%%%%%%%%%%%%%%%%%%%%%%%%%%%%%%%%
We thank Erez Braun, Jacob Rubinstein and Yotam Gil for their help.
We thank Yair Shokef, Ronen Avni and Michael Sheinman for fruitful
discussions. This research was supported in part by the US-Israel
Binational Science Foundation, and by the Yeshaya Horowitz
association through the Center for Complexity Science.

\end{document}